# Spectral focusing of broadband silver electroluminescence in nanoscopic FRET-LEDs

R. P. Puchert, F. Steiner, G. Plechinger, F. Hofmann, I. Caspers, J. Kirschner, P. Nagler, A. Chernikov, C. Schüller, T. Korn, J. Vogelsang, S. Bange, and J. M. Lupton[*)]

Institut für Experimentelle und Angewandte Physik, Universität Regensburg, D-93040 Regensburg, Germany

Few inventions have shaped the world like the incandescent bulb. While Edison used thermal radiation from ohmically heated conductors, some noble metals exhibit "cold" electroluminescence (EL) in percolation films[1,2], tunnel diodes[3], electromigrated nanoparticle aggregates[4,5], optical antennae[6], or scanning-tunnelling microscopy (STM)[7-9]. The origin of this radiation, which is spectrally broad and depends on applied bias, is controversial given the low radiative yields of electronic transitions. Nanoparticle EL is particularly intriguing since it involves localized surface-plasmon resonances with large dipole moments. Such plasmons enable very efficient non-radiative fluorescence resonance energy transfer (FRET) coupling to proximal resonant dipole transitions. We demonstrate nanoscopic FRET-LEDs which exploit the opposite process, energy transfer from silver nanoparticles to exfoliated monolayers of transition-metal dichalcogenides (TMDCs)[10]. In diffraction-limited hotspots showing pronounced photon bunching, broadband silver EL is focused into the narrow excitonic resonance of the atomically thin overlayer. Such devices may offer alternatives to conventional nano-LEDs[11] in on-chip optical interconnects.

[*)] Email: john.lupton@ur.de

A particularly intriguing route to forming electroluminescent metal nanoparticle structures was reported by Dickson *et al.*, exploiting electromigration effects in thin gold and silver films[5]. Such "break junctions" were originally developed to contact single molecules in single-electron transistors[12]. One report even refers to deterministic single-photon emission from single EL hotspots[4]. Although lateral break-junction LEDs are less well-defined and harder to characterise structurally than vertical STM-EL[8], they are easy to make reproducibly without high-vacuum needs and exhibit qualitatively similar spectral features. While there is as yet no comprehensive theory to explain light generation, a model put forward describes how shot noise in single-electron tunnelling between tip and surface leads to the formation of the electronic dipole necessary for light generation[13]. Break-junction LEDs are, in principle, CMOS compatible and could potentially act as light sources for on-chip optical communication[11]. However, this avenue has not been pursued given the spectral heterogeneity between different hotspots and the overall spectral width[5]. As with high-gap GaN LEDs, this problem could potentially be overcome by using spectral converters. Conventional down-converter phosphors rely on simple reabsorption of EL, radiative transfer of energy. This process only works for narrow-band light sources with high luminescence quantum yields. In contrast, non-radiative energy transfer – FRET coupling between the dipole responsible for EL (the donor) and the dipole of a spectral converter (the acceptor) – will give rise to spectral focusing if the radiative quantum yield of the acceptor is greater than that of the donor and FRET occurs prior to non-radiative decay. However, while FRET works well in OLEDs[14], where monolayers are easily integrated, it has proven hard to demonstrate between the surface of an inorganic LED and resonant absorbers, such as a quantum-dot monolayer[15]. The main challenge lies in the fact that FRET necessitates nanometre proximity between the electrically driven donor and the acceptor, which is hard to achieve in conventional LED structures. As we demonstrate, such proximity arises naturally by transferring a TMDC monolayer onto an electroluminescent silver break-junction structure. The emissive species in the silver break junction couples resonantly with the TMDC exciton, changing the spectrum of light emitted through the silver film.

Figure 1a sketches the device in top and side view. A 20nm-thin silver square is evaporated on a glass substrate and contacted by a thicker film, lithographically structured into two electrode leads. A d.c. voltage is applied and increased slowly. Once electromigration sets in, the current drops suddenly and EL spots appear at the centre of the film[5]. To prevent continued electromigration, the device is subsequently driven by a radio-frequency alternating current. Using a stamp-transfer technique[16], a single-layer TMDC crystal is placed on the silver film prior to electromigration. Panel (b) shows a CMOS microscope image of the device post electromigration, illuminated by a green LED and viewed from beneath through the thin silver film. The contours of the TMDC flake are indicated in yellow. EL is seen from individual hotspots. Panel (c) compares EL spectra from the red and blue spots in (b) – with and without a $MoS_2$ flake atop the silver. The flake focuses the EL spectrum to its narrow excitonic resonance, so that both higher- and lower-energy components of the silver EL become less significant. In this example, the $MoS_2$ spot appears 35 times brighter than the silver. This difference presumably arises because of the higher conductivity and thus higher hotspot bias beneath the flake than besides it, so we attach little significance to this observation. Since light does not pass through the TMDC overlayer, $MoS_2$ does not simply serve as a spectral filter. This approach of generating EL from TMDC monolayers therefore differs fundamentally from conventional p-n device structures with direct carrier injection[17-19]. EL occurs from diffraction-limited hotspots as seen in the close-up image in Fig. 1d, where the point-spread function has a width of 360 nm. The brightness of the spot increases exponentially with applied bias, plotted in Fig. 1e, demonstrating the predominance of a tunnelling mechanism in the electromigrated silver film[1]. Further insight into the origin of EL is gained from photon statistics, the second-order autocorrelation function of the EL $g^{(2)}(\Delta\tau)$.

This is determined by passing the luminescence through a beam splitter and recording the coincidence rate on two detector channels as a function of delay time $\Delta\tau$ between the first ("start") and the second ("stop") photon detected. To avoid the emergence of a characteristic timescale in photon emission due to the a.c. excitation, we drive the device at low d.c. bias, below that used for electromigration. This condition reduces overall EL stability since electromigration continues slowly during EL, but otherwise leads to identical emission characteristics. Clear signatures of photon bunching ( $g^{(2)}(\Delta\tau)>1$ ) from the nanoscale light source are resolved in Fig. 1f, yielding a characteristic time constant in emission of ~2 ns. We assign this bunching to the charging of silver nanoparticles formed by the initial electromigration. These particles effectively constitute a capacitor structure, with a characteristic RC time[20]. Once the bias has built up across the junction, the particles can discharge by a tunnelling current. Tunnelling-induced EL therefore contains information on electronic transport in the statistics of the emitted photons.

FRET from excitonic donors to metallic acceptors has been studied extensively and always results in efficient long-range quenching of fluorescence since the acceptor dipole of the plasmon resonance is so large[21-23]. A few reports have addressed the possibility of FRET occurring from molecules or quantum dots to two-dimensional acceptors such as graphene or TMDC monolayers, and have revealed a weaker distance dependence – i.e. stronger coupling – than in conventional point-dipole Förster transfer due to the large oscillator strength and delocalization of the electronic excitation[24-26]. Here, we use the metal to *generate* light – as the donor – but clearly the same arguments regarding non-radiative dipole-dipole coupling still hold[22]. This reversal of donor-acceptor roles in the metal-fluorophore structure becomes possible because of the extremely short (sub-150 fs) radiative exciton lifetime of TMDC monolayers in conjunction with the giant oscillator strength of the delocalized excitation[27].

Because FRET is expected to predominantly create TMDC excitons within the light cone, these can very rapidly recombine radiatively[27] prior to undergoing diffusion or quenching.

We examine the spectral focusing effect by comparing a range of raw data. Figure 2 plots EL spectra for different combinations of materials. Panel (a) shows a typical hotspot spectrum from a bare silver film, which is broad and mostly featureless. To illustrate the scatter between individual hotspots, we plot 150 such normalised spectra on a false-colour scale in panel (b), with the abscissa indicating measurement index number. The spectra scatter widely between hotspots, spanning the entire visible region into the IR. The right-hand panel shows the sum of all normalised spectra, providing a measure of spectral average. Figure 2c,d gives the corresponding spectral sampling for $MoS_2$ FRET-LEDs, again with no selection or manipulation of the data carried out. The EL now appears focused into a narrow band around 720 nm, close to the first excitonic transition of $MoS_2$. For comparison, the photoluminescence (PL) spectrum of a flake on a $SiO_2$/Si substrate is shown with a typical EL spectrum in panel (c). The EL spectra are generally slightly red shifted from the PL spectra. Small variations in peak position and in background are apparent in the scatter plot in (d), but overall most of the EL clearly originates from the TMDC overlayer. On the right-hand side of panel (d) we also plot the imaginary part of the dielectric function $\varepsilon''$ of $MoS_2$, extracted from reflectometry measurements of a monolayer on an $SiO_2$/Si substrate as described previously[28-29]. These data provide a measure of the underlying excitonic absorption resonances. For FRET to occur, donor (silver) emission and acceptor ($MoS_2$) absorption must overlap, which is evidently the case. As discussed below, the small shift between the EL spectra and the PL as well as the resonances seen in $\varepsilon''$ most likely arise from strain in the monolayer as a result of electromigration. The spectral scatter in the hotspot EL [panel (d)] appears slightly larger on the long-wavelength side, with an apparent cut-off at short wavelengths. This effect is more pronounced for the $WS_2$ (Fig. 2e,f) and $WSe_2$ (Fig. 2g,h) overlayers and is apparent as a shadow in the false-colour plots, which roughly coincides with the lowest-energy absorption resonance in $\varepsilon''$.

Most EL hotspots show coupling to the excitonic resonances of the two-dimensional semiconductors, so that the hotspot spectra are tuned by the choice of overlayer. Spectral focusing apparently works best with the $MoS_2$ overlayer. This material has the broadest absorption feature which overlaps most with the silver EL spectrum, and the smallest splitting between A and B excitons (the two peaks in the green spectrum in Fig. 2d[28]) so that both resonances overlap with the donor. FRET efficiency can also be probed by considering absolute intensities, which are summarized in the histogram in Fig. 2i for a total of 477 hotspots. In the spectral measurements, we optimized the brightness and stability of the single hotspots by tuning the driving voltage between 2-4 V. While it is not possible to compare one and the same hotspot with and without the overlayer, the distribution of intensities contains valuable information. The scatter is narrowest for $MoS_2$, implying that coupling of the silver donor to the acceptor increases overall hotspot stability and reproducibility. This improvement may arise because the luminescence quantum yield of $MoS_2$ is expected to be much more uniform than that of silver nanoparticles; FRET prior to the onset of non-radiative decay in the silver will therefore homogenise the observed hotspot brightness.

The stamping technique only works on smooth layers prior to electromigration, which can be followed by non-contact atomic-force microscopy (AFM). Figure 3 shows AFM topology images of a silver film prior to [panel (a)] and after [panel (b)] electromigration, when the break-junction region becomes visible. Panel (c) shows an electromigrated $MoS_2$ sample. The flake edge is visible as a change in contrast in the image and is indicated by yellow arrows. Interestingly, the features of the break junction beneath the TMDC layer appear smoothed and lower in contrast. This effect is clearly seen by comparing line scans along the red and black arrows in the image, plotted in panel (e): the TMDC flake appears propped up like a tent over the nanoparticles. This effect is particularly apparent in the phase-contrast AFM image of the same region [panel (d)]. Electromigration should therefore lead to tensile, predominantly biaxial strain in the TMDC layer, which shifts the spectrum to longer wavelengths[30].

Strain can be discerned by spatially resolving the PL spectra of the TMDC layer. Figure 4a compares an EL spectrum to PL spectra of $WS_2$ stamped on a $SiO_2$/Si wafer, on a silver film, and on the break-junction region. The former show only one PL peak (Peak I). After electromigration, an additional PL peak is resolved in the break junction region (Peak II), but only the lower-energy feature appears in EL. Figure 4b shows a microscope image of an operating device, with the contour of the flake indicated in white and the break-junction region illuminated by the hotspots. Panels (c),(d) plot the same region under a fluorescence microscope, colour coding the peak PL wavelength. The entire flake shows up if PL is selected between 605-645 nm, i.e. Peak I in panel (a). However, for a cut-off wavelength of >650 nm (Peak II), only the region in the vicinity of the break junction is detected. The appearance of a new, red-shifted peak in the break junction is consistent with straining of the TMDC layer by the nanoparticle formation seen in the AFM images in Fig. 3. EL is therefore dominated by the low-energy feature of the PL spectrum corresponding to a strained 2D crystal. Given the finite spatial resolution of the fluorescence microscope, however, unstrained regions will also be recorded in the break-junction PL, implying that the EL-active areas of the layer must be smaller than the diffraction limit. Strain also explains the slight systematic red-shift of the short-wavelength EL cut-off in Figs. 2d,f,h compared to the fundamental exciton transition.

TMDC-monolayer-based nanoparticle EL could be appealing for on-chip optical communication because, in contrast to carbon-nanotube LEDs[11], it combines potentially high integration density with tuneable emission in a CMOS-compatible approach. In addition, such nano-LEDs could find use in sub-diffraction transmission microscopy[31] of two-dimensional crystals, without the limitations and complexities of conventional scanning nearfield probes[32].

## Methods

### *Sample preparation*

To ensure reproducible break-junction formation, the devices were designed using a lithographic process. First, square regions of $100 \times 100 \mu m^2$ were defined with a mask aligner in a spin-coated double photoresist (AR300-80, AR-P5350, Allresist GmbH) and subsequently developed (AR300-35, Allresist GmbH). After evaporation of a 20nm thick silver layer, the active structure was formed by acetone lift-off of the residual photoresist. In a second photolithography step, the electrode leads were defined. These consist of a 150nm thick silver film. Because of the homogeneity of the process, the break junction always forms in the centre of the thin silver layer. The TMDC monolayers were mechanically exfoliated from bulk crystals onto a polydimethylsiloxane (PDMS) film. Subsequently, they were stamped onto the silver film by a dry deterministic transfer based on the viscoelasticity of the PDMS substrate[17]. As this procedure is performed under an optical microscope, micronscale control over the position of the flake on the silver film is achieved.

### *Optical spectroscopy*

Electromigration can be carried out under inert conditions such as under vacuum or under dry nitrogen flow. Typically, the sample was inserted into a cryostat (Supertran ST-500, Janis Research) and connected to a voltage source (Keithley 2401). The d.c. voltage was increased slowly up to the point at which the resistance of the silver film rises rapidly. At this voltage, the gap in the film, the break junction, is formed. For optical measurements, the electromigrated sample was connected to an amplified high-frequency source (Amplifier

ZHL-2-S, Mini Circuits; HF generator 83732A, Agilent Technologies). The frequency was set to 100MHz and the voltage measured at the sample was set to between 2 and 4V, optimized for the hotspot brightness and stability. The EL was collected by a microscope objective (SLCPlanFl 40×, Olympus) through the glass substrate. Images of the sample, illuminated by a green LED, and the EL hotspots were recorded with an sCMOS camera (Orca-Flash 4.0, Hamamatsu) and the spectra were measured with a spectrometer and a CCD camera (Acton SP2300, Pixis 400, Princeton Instruments). We confirmed that the CCD camera and the photon counting technique (see below) yielded approximately the same overall hotspot intensities (photon counts) of order 100 kHz.

PL spectra were measured in a microscope setup, where the light of a c.w. frequency-doubled solid-state laser operating at a wavelength of 532 nm is focussed onto the sample to a spot size of ~1μm. The PL from the TMDC monolayer was then analysed with a spectrometer (Acton SP2750, Princeton Instruments) with a 150 lines per mm grating and a Peltier-cooled CCD chip. For spatial PL maps, spectra were recorded for every pixel (0.5μm step size) and the resulting PL peaks were fitted with Gaussian functions in a predefined spectral range. The centre wavelength of the fit function is then displayed in the colour-coded map in Figure 4.

The photon correlation measurements were performed using an oil-immersion objective (UPLSAPO 60×, 1.35 numerical aperture, Olympus) where the light is detected by two single-photon detectors (PicoQuant t-spad 20) in a Hanbury Brown-Twiss setup with a 50:50-beam splitter (Thorlabs BS013) in between both detectors. The correlation data were analyzed with commercial software (PicoQuant SymPhoTime64). The data in Fig. 1(f) were fitted with a single-exponential function $g^{(2)}(\Delta\tau) = A \cdot e^{-\frac{\Delta\tau}{\tau}}$ with a time constant τ=2ns. Since avalanche photodiodes themselves generate luminescence in the avalanche process, autocorrelation measurements typically show features of these low-energy secondary fluorescence photons detected at a fixed time delay by the opposite detector. These photons are seen in the autocorrelation in Fig. 1f in the afterglow peak at 8ns.

*White-light reflectometry*

The imaginary components of the complex dielectric function ε´´(λ), which is proportional to the optical absorption of the TMDC monolayers studied, are shown in Fig. 2. We performed white-light reflectometry on TMDC layers deposited on $SiO_2$/Si substrates since silver substrates are poorly suited for either transmission or reflection microscopy. For white-light reflectance measurements, the output of an incandescent lamp was focussed onto a pin hole and subsequently collimated with an achromatic lens. The light beam was coupled into a microscope setup with a 100× objective lens, so that the white light could be focussed down to a spot size of approximately 4 μm on the sample surface. The reflected light was collected with the same objective lens and coupled into a spectrometer, where it was detected with a Peltier-cooled CCD. Subsequent measurements on the TMDC flakes and on the bare substrate were recorded, and the reflectance contrast was determined from the reflected light intensities as $\Delta R(\lambda) = (R_{TMDC} - R_{substrate})/R_{substrate}$. The dielectric response is extracted from the measured reflectance contrast spectra by a constrained Kramers-Kronig approach[28-29], including multi-Lorentzian parameterization of the complex dielectric function and by taking into account the presence of the substrate via a transfer-matrix method to match the experiment with properly adjusted parameters. To analyse the data, a physically meaningful number of Lorentzians was used to account for only the most pronounced optical transitions in the spectral range of interest, including the A and B exciton resonances.

*AFM measurements*

AFM images were taken with silicon probes (Nanosensors PPP-NCHR and SSS-NCHR) using a Park Systems XE-100 in "non-contact" amplitude modulation mode under ambient conditions in a vibration-isolated acoustic enclosure.

**Acknowledgements**

We are indebted to the DFG for offering collaborative funding through the SFB 689 and GRK 1570. JML is grateful for financial support from the European Research Council through the Starting Grant MolMesON (#305020). AC acknowledges funding from the DFG through the Emmy Noether Programme (CH 1672/1-1).

**Figure captions**

**Figure 1. Nanoscopic FRET-LEDs from electromigrated silver nanoparticles covered with an exfoliated TMDC monolayer.** a) Cartoon of the device structure. An evaporated 20nm-thin silver square is contacted by thicker electrode leads. Electromigration gives rise to the formation of a break junction at the centre of the thin square, where EL occurs. The TMDC layer is placed on the smooth silver film prior to electromigration (top left), so that the nanoparticle junction forms beneath the layer (top right). Excitation energy is transferred through resonant dipole-dipole coupling (FRET) from the electroluminescent metal hotspot to the TMDC overlayer (bottom). EL is detected from beneath. b) Microscope image of a break-junction structure with a $MoS_2$ flake (yellow), illuminated by an external LED. Distinct EL hotspots are visible. c) Normalised EL spectra of the two single hotspots marked in b) showing spectral funnelling. The silver spectrum (blue) appears narrowed by the $MoS_2$ layer (red). d) Microscope image of a diffraction-limited EL hotspot. e) Brightness-voltage characteristic of a single EL hotspot with an exponential fit to the data. f) Photon correlation of a single hotspot showing photon bunching on the timescale of 2 ns, which presumably corresponds to the RC time constant of the nanoparticle aggregate. The red curve shows an exponential fit to the correlation with the Poisson limit of classical photon emission ( $g^{(2)}(\Delta\tau)$ =1) marked by a grey line. The peak at 8 ns in the autocorrelation, marked in green, originates from the secondary photon emission (afterglow) of the avalanche photodiodes (see *Methods*).

**Figure 2. Spectral focusing of nanoscopic silver FRET-LEDs with different overlayers.** The first column shows a representative single-hotspot spectrum and the second column indicates the scatter of normalised single-hotspot spectra, plotted on a false-colour intensity scale. The abscissa refers to an arbitrary sample number and contains no information. The third column shows the average of the normalised spectra and the imaginary component of the TMDC monolayer dielectric function $\varepsilon''$, which largely determines absorption and indicates the dominant excitonic resonances. $\varepsilon''$ was obtained by reflectometry of monolayer samples on $SiO_2/Si$ substrates. a-b) Bare silver film, and overlayers of $MoS_2$ (c-d), $WS_2$ (e-f) and $WSe_2$ (g-h). PL spectra of TMDC flakes on $SiO_2/Si$ are also shown for comparison. i) Histograms of the absolute hotspot brightness. Note that the driving voltage was adjusted between 2-4 V to provide comparable brightness for the different structures.

**Figure 3. Non-contact atomic-force microscopy of the break-junction region.** a) Topography of a smooth silver film, an electromigrated silver film (b) and an electromigrated silver film with an $MoS_2$ layer transferred prior to electromigration (c). The edge of the TMDC flake is marked by yellow arrows. d) Image of the same region as in (c) in AFM phase, clearly revealing the TMDC flake domain. e) Topography line scans of the regions marked in panel (c) showing a smoothing of topographic features when imaged through the TMDC overlayer.

**Figure 4. Change of $WS_2$ PL spectrum upon electromigration due to straining.** a) PL spectrum of a flake on a $SiO_2$ wafer, on a silver film evaporated on glass, and over a break junction. While the flat metal does not affect emission, two peaks are seen in the break-junction region. EL appears to originate from the lower-energy Peak II, which is associated with straining of the TMDC layer. b) Microscope image of the break-junction region illuminated by EL hotspots. The position of the flake is indicated in white. c) Spatial map of the peak wavelength of Peak I for the same region given in b), shown for all emission wavelengths between 605 nm and 645 nm. The entire $WS_2$ flake is visible. d) As in (c) but for spectral selection of the PL microscope image for wavelengths >650 nm (Peak II). Only the region of the flake in the vicinity of the break junction is visible.

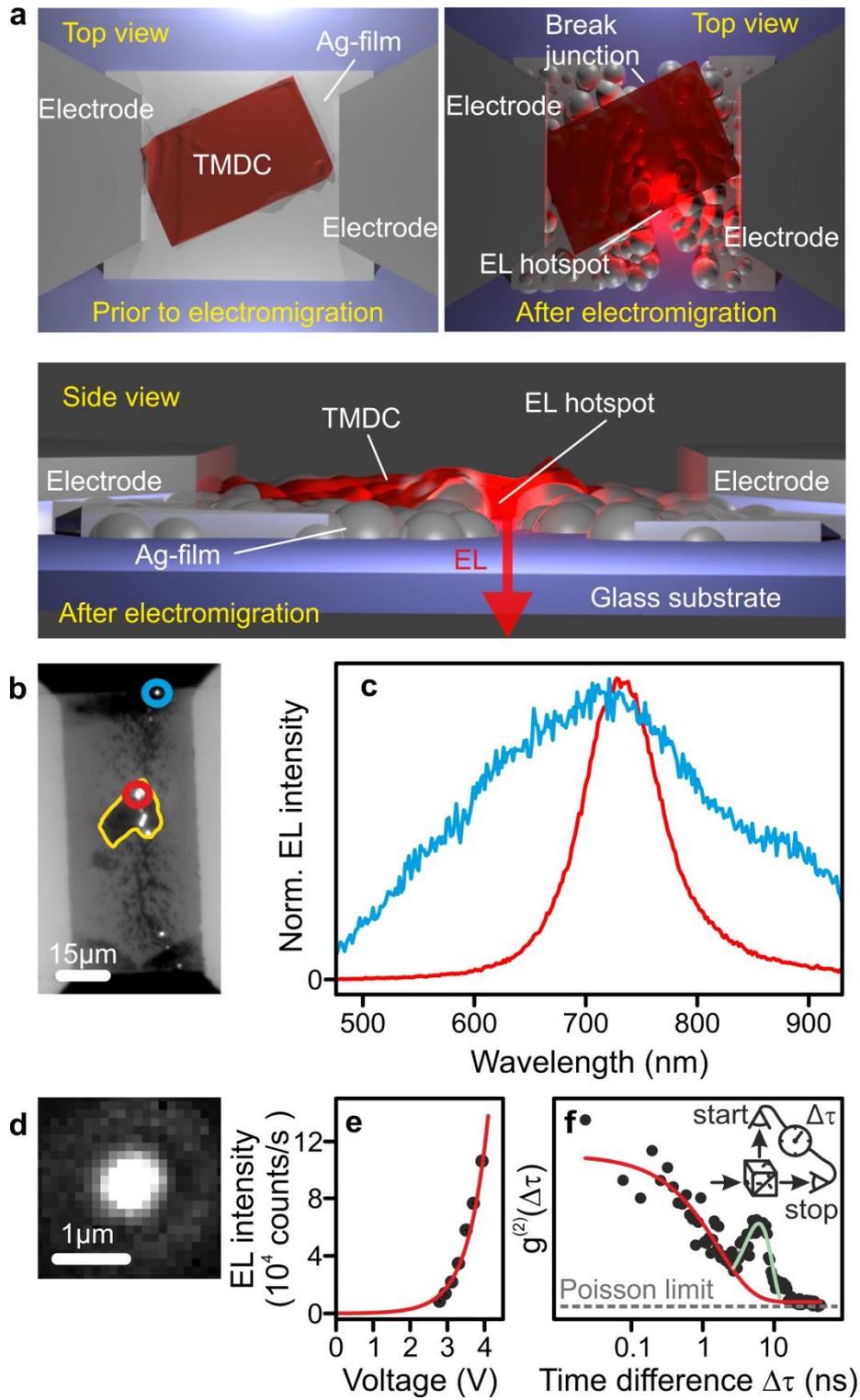

Figure 1.

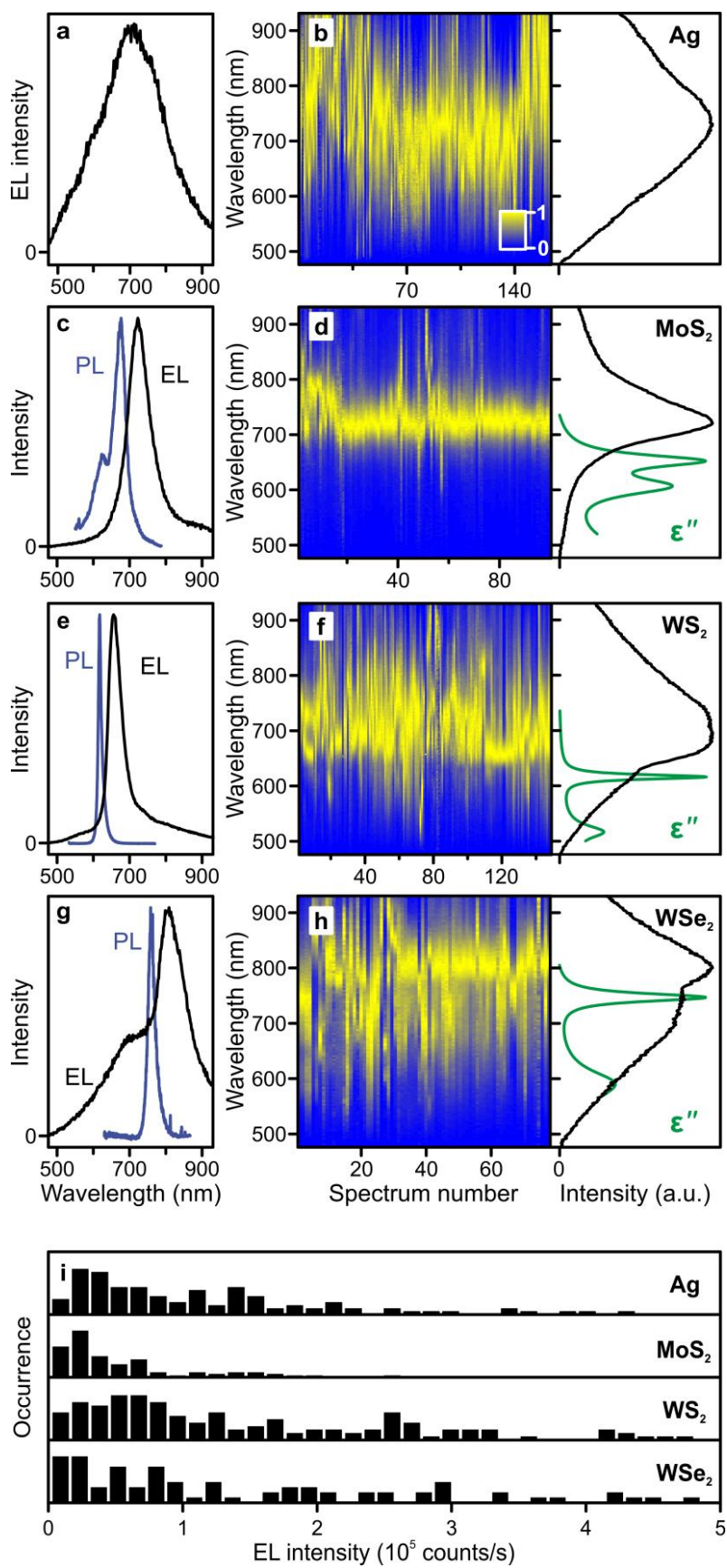

Figure 2.

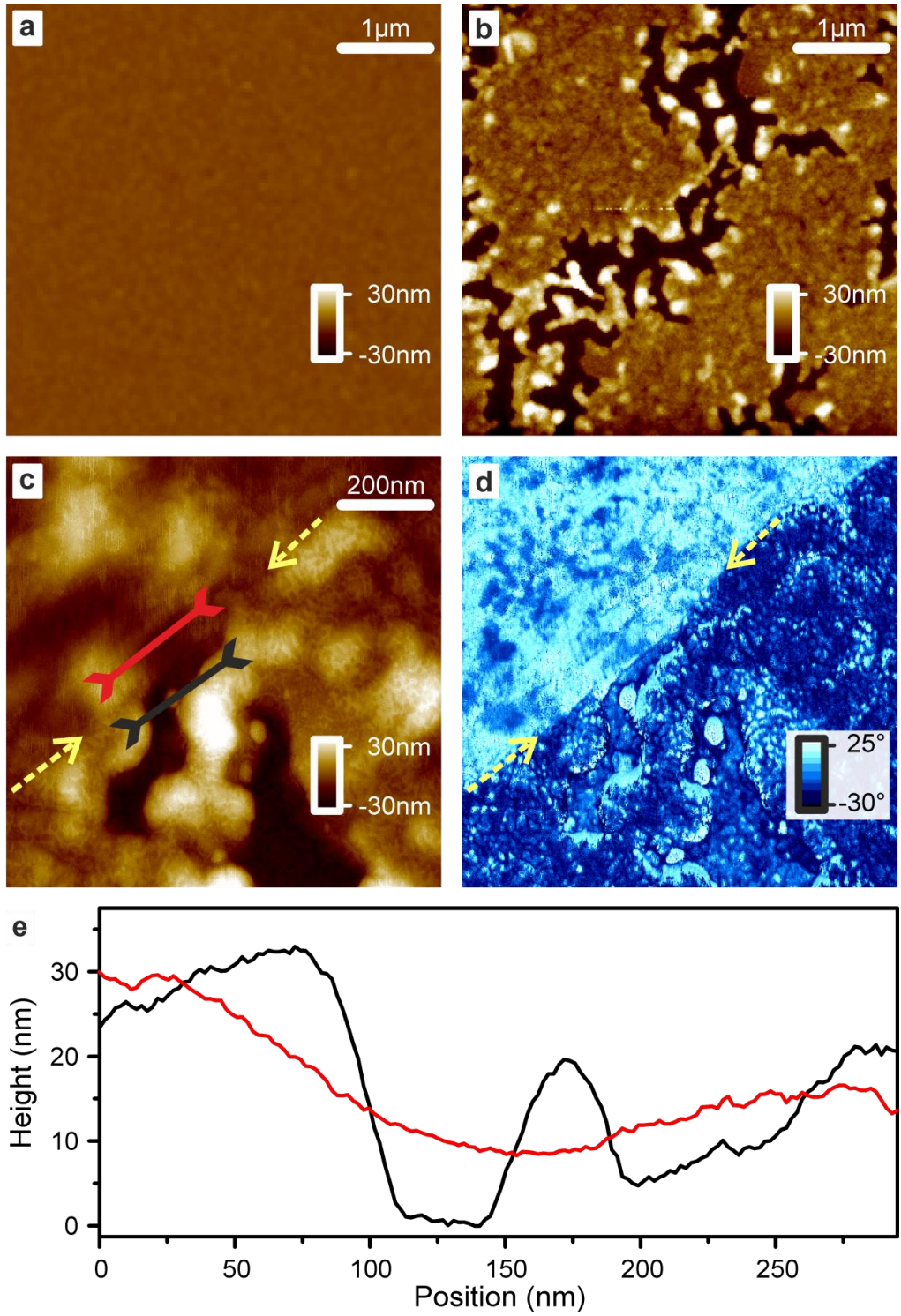

Figure 3.

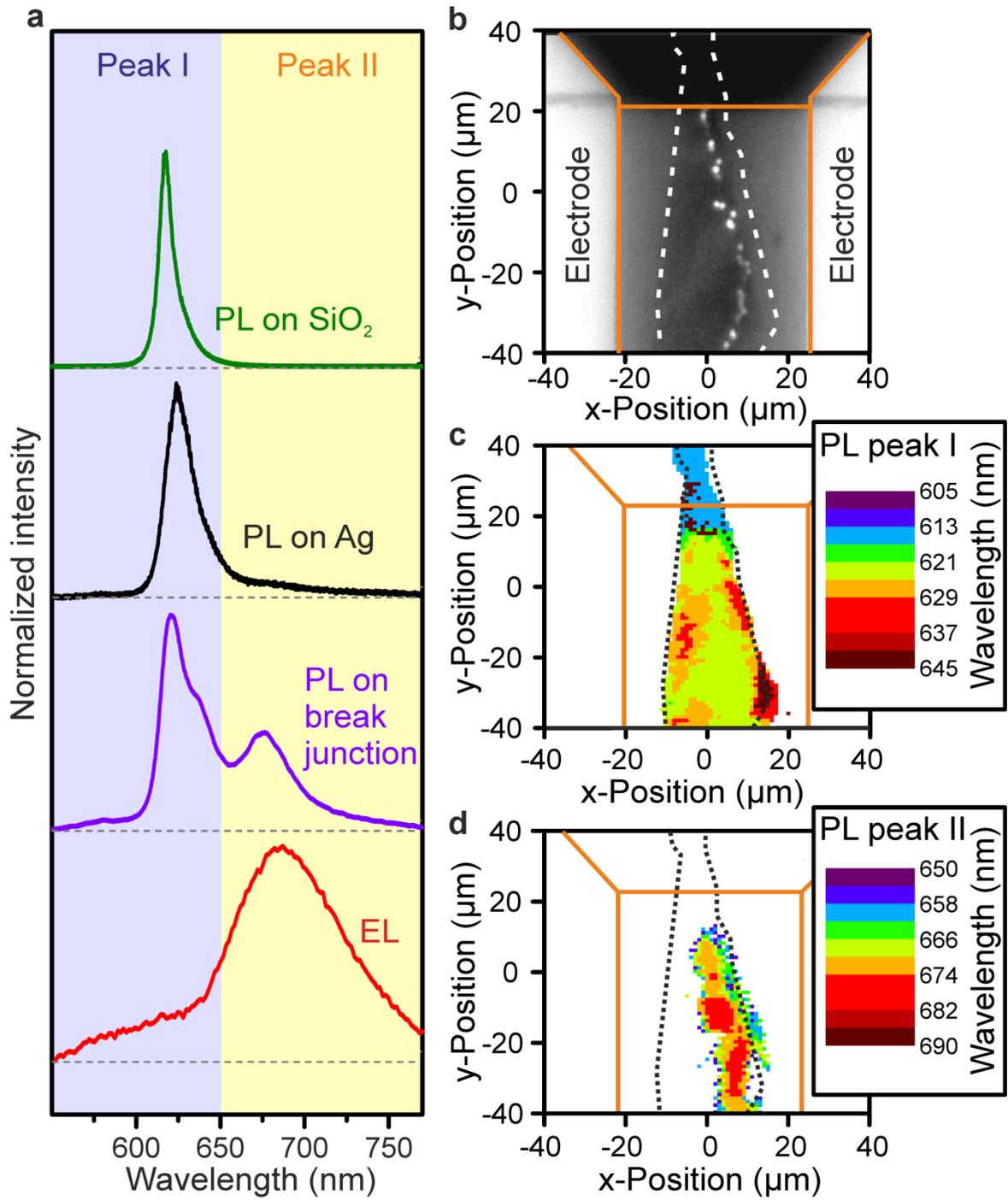

Figure 4.